\documentclass[12pt,a4paper,english]{paper}
\usepackage[T1]{fontenc}
\usepackage[latin1]{inputenc}
\usepackage{graphicx}

\makeatletter

\newcommand{\lyxline}[1]{
  {#1 \vspace{1ex} \hrule width \columnwidth \vspace{1ex}}
}

\newcommand{\lyxdot}{.}

\newcommand{\lyxaddress}[1]{
\par {\raggedright #1
\vspace{1.4em}
\noindent\par}
}

\usepackage{babel}
\makeatother
\begin{document}

\title{Events with Isolated Leptons and Missing Transverse Momentum in \emph{ep}
Collisions at HERA}

\author{Gerhard Brandt%
\thanks{email: \texttt{gbrandt@mail.desy.de}%
}}

\maketitle

\lyxaddress{Physical Institute, Philosophenweg 12, 69120 Heidelberg}

\begin{abstract}
The analysis of events with isolated leptons and missing transverse
momentum in the H1 experiment is discussed for the electron, muon
and tau channels. In the Standard Model (SM) framework, production
of real $W$-bosons gives rise to such topologies. Contributions to
the background are dominated by QCD processes. An excess of observed
signal over background presents a chance of the discovery of new physics.
The results using the HERA 1994-2006 data set corresponding to 341
pb$^{-1}$ are presented While the $e^{-}p$ sample shows good agreement
between data and SM expectation, in $e^{+}p$ collisions an excess
over the SM expectation with 3.4$\sigma$ significance is observed
at high hadronic transverse momentum. 
\end{abstract}
\thispagestyle{empty}

\section{Introduction}

Events with high-$P_{T}$ isolated leptons and missing transverse
momentum are sensitive to physics beyond the Standard Model (BSM).
Both H1 and ZEUS have previously published searches for events with
isolated electrons and muons \cite{IsoLep94,IsoLep98,IsoLepZEUS00,IsoLep03,IsoLepZEUS03}
and $\tau-$leptons \cite{TauZEUS04,Tau06}. The observation of such
events in $ep$ collisions at HERA has created considerable interest
in the high-energy physics community (many references can be found
in \cite{IsoLep98}) and is continously updated as new data become
available \cite{IsoLepICHEP06,IsoLepZEUSICHEP06,TauPrelim}. Section
2 reviews the experimental conditions (collider and detector) of the
search, section 3 presents the most important physics processes involved,
section 4 presents the search for $e/\mu+P_{T}^{\mbox{miss}}$ events,
section 5 the search for $\tau+P_{T}^{\mbox{miss}}$ and section 6
discusses the results and possible interpretations.

\section{Experimental Conditions}

HERA operating at DESY in Hamburg is the worlds only $ep$ collider.
It collides electrons or positrons%
\footnote{In this article, the term \emph{electron} refers to both electrons
and positrons unless explicitly stated.%
} with protons at a center-of-mass energy $\sqrt{s}=320$ GeV for most
of its operating time. HERA data taking is divided in two main running
periods, HERA-1 in the years 1994 to 2000, and HERA-2 in the years
2000 to 2007. During HERA-1, H1 collected a data set corresponding
to 118 pb$^{-1}$ of data, dominated by about 90\% by $e^{+}p$ data.
For the HERA-2 period, the luminosity has been upgraded and spin rotators
installed to provide a longitudinally polarised electron beam. For
the results discussed here about 250 pb$^{-1}$ of HERA-2 data have
been available to H1. This includes the full $e^{-}p$ data sample
which corresponds to a total integrated luminosity of 182 pb$^{-1}$.
Data taking continues with $e^{+}p$ running until the end of HERA
operation, which is scheduled for June 2007.

\begin{figure}[tbh]
\textbf{a)} \includegraphics[height=7cm]{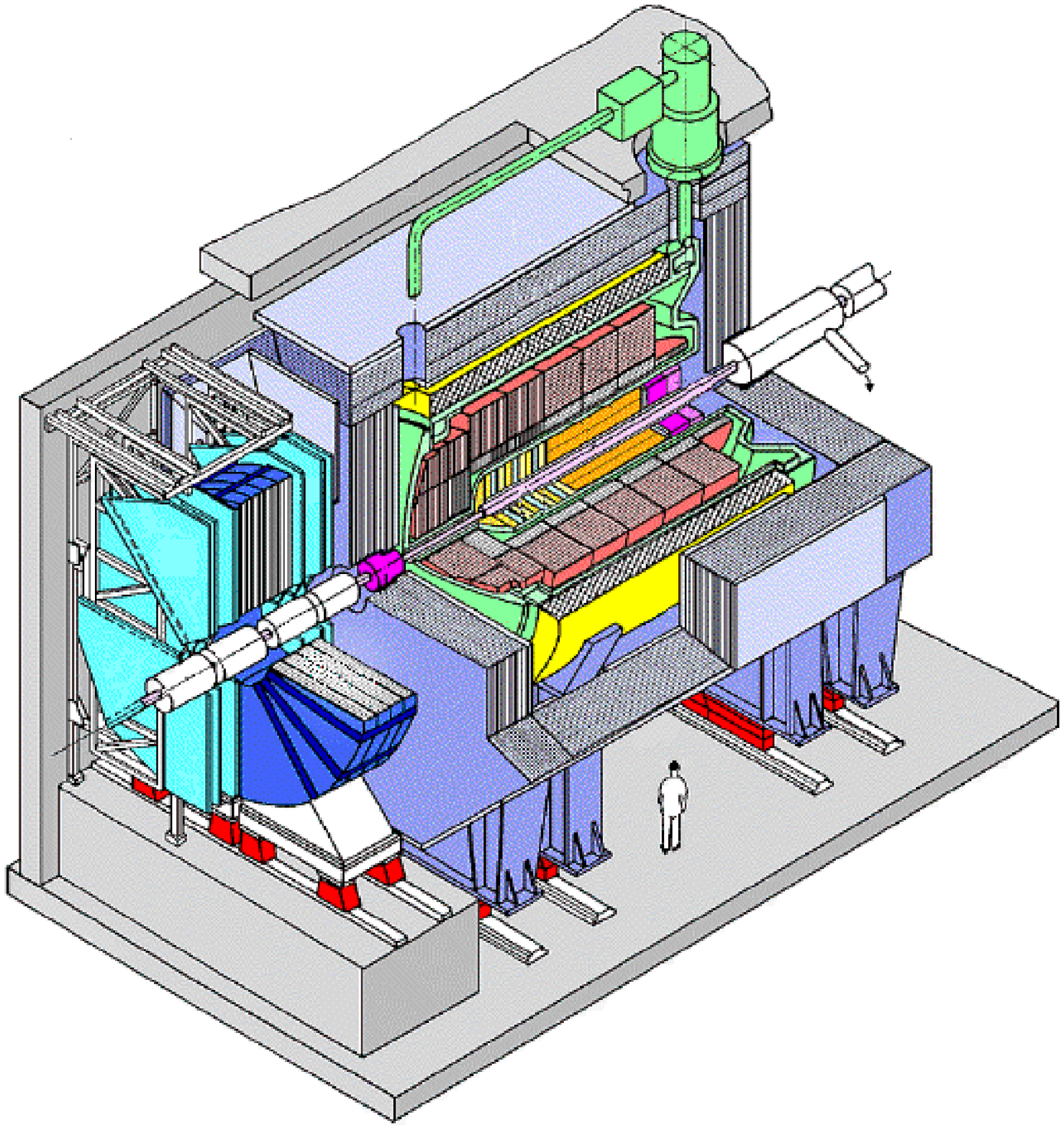} \textbf{b)}
\includegraphics[height=6cm]{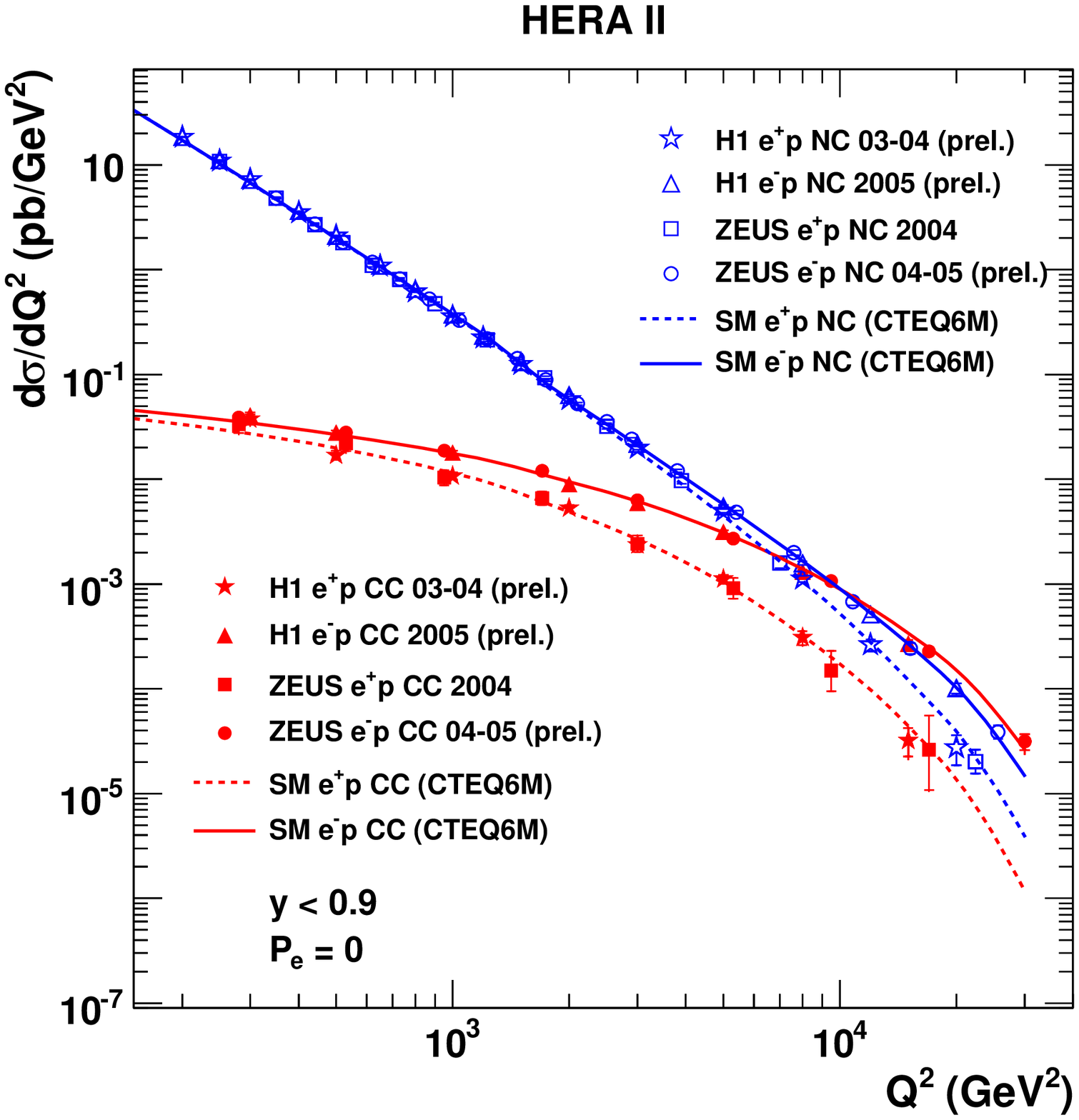}

\caption{\label{fig:H1}\textbf{a)} The H1 detector at HERA \cite{H1Det}.
Electrons enter from the left side, and protons from the right side;
\textbf{b)} A HERA {}``textbook'' measurement: Inclusive unpolarised
cross sections of NC and CC DIS processes in $ep$ collisions measured
by the H1 and ZEUS experiments (taken from \cite{NCCC06}). The unification
of electroweak forces is visible for $Q^{2}>1000$ GeV.}
\end{figure}

Two collider experiments, H1 and ZEUS, operate at HERA. This article
focusses on H1, described in detail in \cite{H1Det}. A drawing of
the H1 detector is shown in figure \ref{fig:H1}a. It has almost $4\pi$
angular coverage to detect missing transverse momentum and the subdetectors
are arranged for excellent lepton identification. Track information
is provided by central jet chambers, a forward tracker and silicon
vertexing detectors. The trackers in a solenoid magnet field allow
momentum measurement of charged particles. Energy measurement for
electromagnetic and hadronic showers is provided by a LAr (liquid
argon) calorimeter in the central and forward (proton beam direction)
region, and a lead-fibre spaghetti calorimeter ({}``SpaCal'') in
the backward direction. For muon detection, the iron flux return yoke
of the magnet is instrumented with limited streamer tubes to measure
escaping muons. In the forward direction, there is an additional forward
muon detector installed.

To compare recorded data to SM predictions, the relevant physics processes
are modelled by Monte Carlo generators. The detector response is simulated
by H1 software based on GEANT 3 \cite{GEANT}. The simulated events
are then processed through the complete reconstruction and analysis
chain, identical to the data treatment.

\section{Physics in \emph{ep} Collisions}

\begin{figure}
\begin{centering}\textbf{a)} \includegraphics[height=4cm]{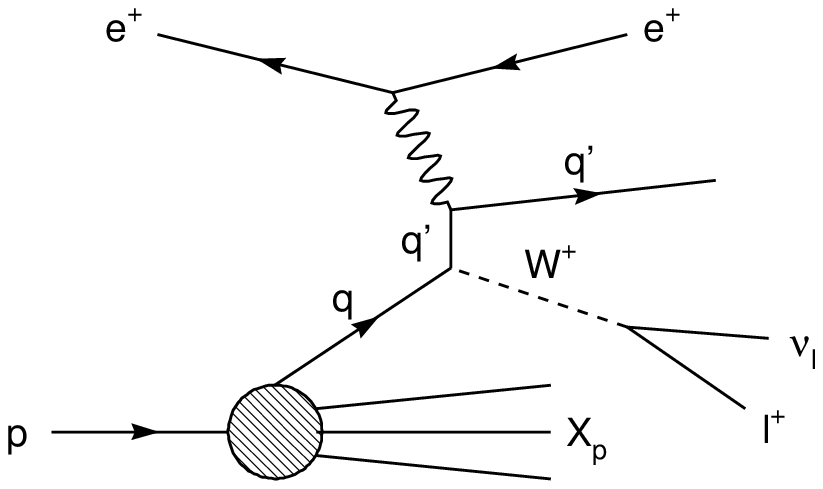}
\textbf{b)} \includegraphics[height=4cm]{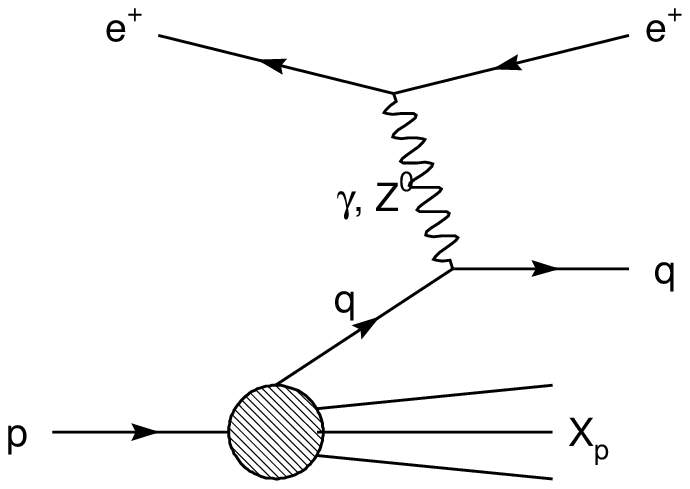} \par\end{centering}

\begin{centering}\textbf{c)} \includegraphics[height=4cm]{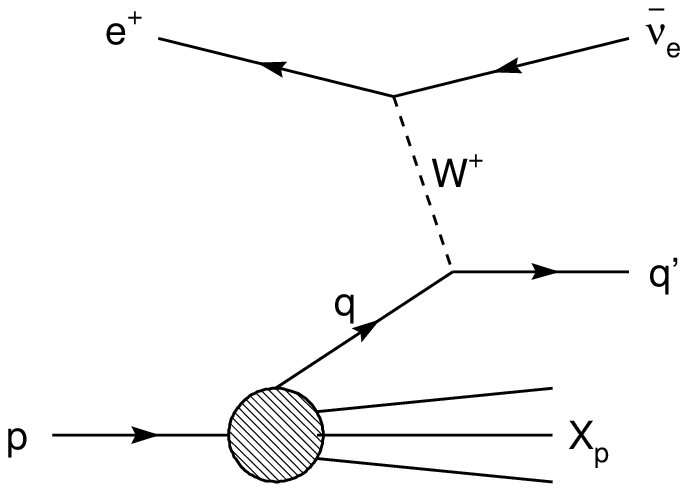}
\textbf{d)} \includegraphics[height=4cm]{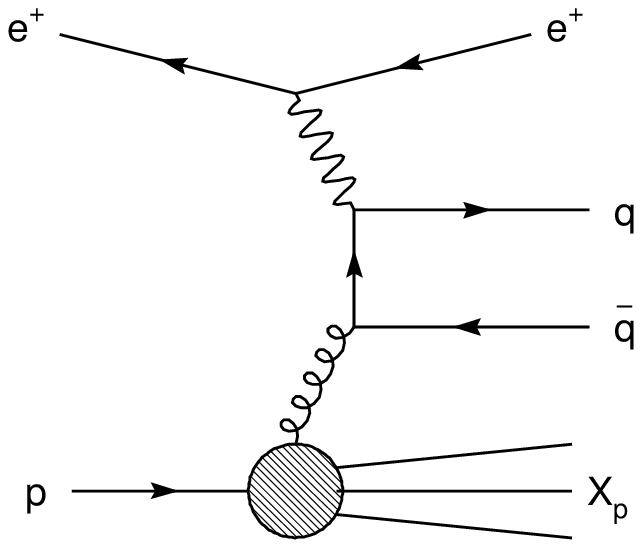} \par\end{centering}

\begin{centering}\textbf{e)} \includegraphics[height=4cm]{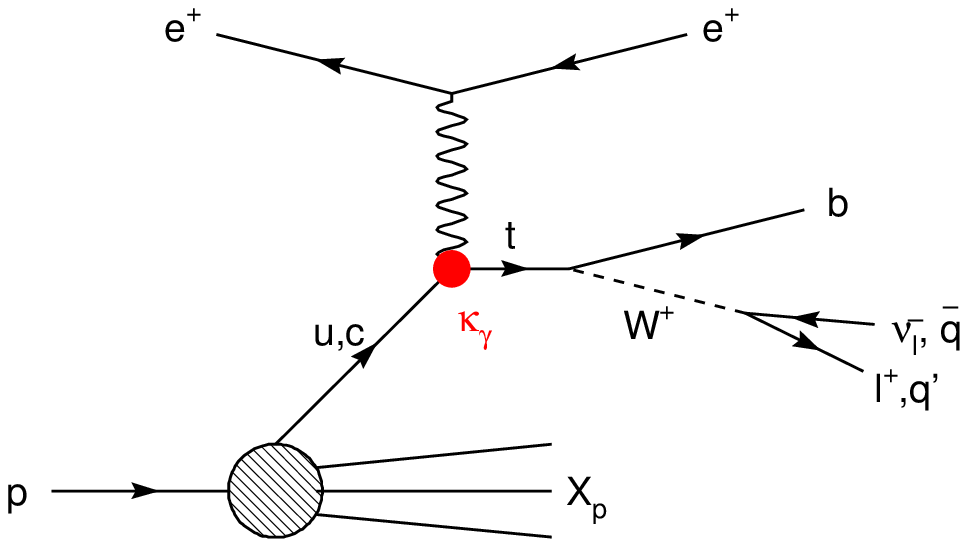}\par\end{centering}

\caption{\label{fig:WProdDiag}Physics processes at HERA contributing to $l+P_{T}^{\mbox{miss}}$
topologies as signal (a and e) or background due to mismeasurements
(b,c,d): \textbf{a)} Production of real $W$-Bosons in the Standard
Model; \textbf{b)} Neutral current; \textbf{c)} Charged current; \textbf{d)}
Photoproduction of jets by boson-gluon-fusion; \textbf{e)} Anomalous
production of single \emph{top} quarks via FCNC, the template used
for possible BSM contributions.}
\end{figure}

\begin{figure}
\begin{centering}\includegraphics[width=12cm]{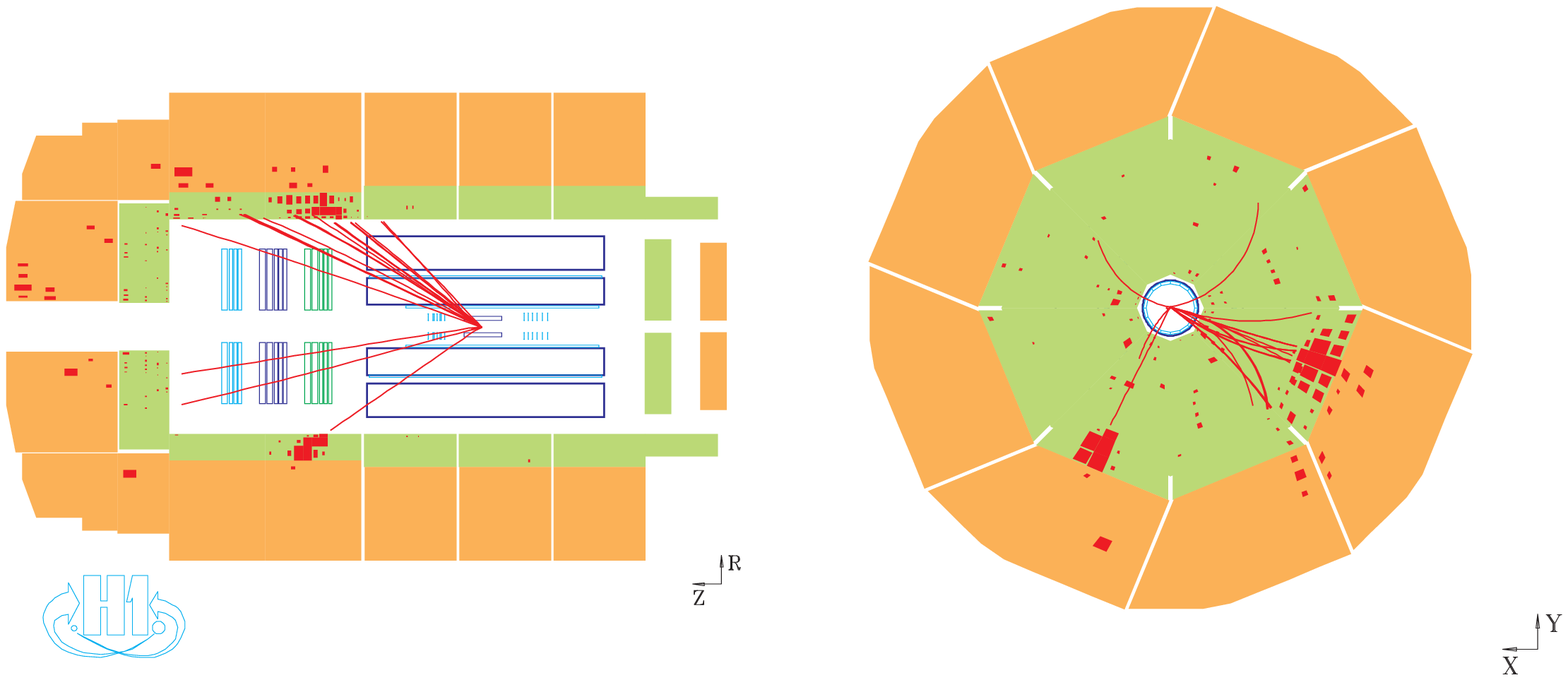}\par\end{centering}

\caption{\label{fig:ElecEvent}Event display for an event with an isolated
electron of $P_{T}^{e}=37$GeV, missing transverse momentum of $P_{T}^{^{\mbox{miss}}}=44$GeV
and a hadronic system of $P_{T}^{X}=29$ GeV, observed in HERA-2 $e^{+}p$
data. The Standard Model expectation for events with this topology
is low.}
\end{figure}

In the Standard Model events with isolated leptons and missing energy
are mainly expected from real $W$ production with subsequent leptonic
decay, \[
ep\rightarrow eW^{\pm}(\hookrightarrow l\nu)X,\]
where X denotes the hadronic final state (HFS). This is composed of
the scattered quark jet and the proton remnant. The main contribution
to this process is radiation of a $W$ boson off the scattered quark
line, shown in figure \ref{fig:WProdDiag}a. This and other contributing
processes are simulated by the EPVEC generator \cite{EPVEC}. The
total cross section is about $1$ pb. With a typical detection efficiency
of 40\% for the total leptonic branching ratio this means about 100
events are expected in an $ep$ data sample of 250 pb$^{-1}$. For
SM $W$ production, the transverse momentum of the HFS $P_{T}^{X}$
is predicted to be well below 25 GeV. At high $P_{T}^{X}$ the SM
prediction is low. Looking at its $P_{T}^{X}$ provides a measure
for how untypical, or interesting, an event is. An event with a $P_{T}^{X}$
of 29 GeV is shown in figure \ref{fig:ElecEvent}. 

Several physics processes contribute to the background. The most prominent
physics processes measured at HERA are neutral current (NC) $ep\rightarrow eX$
and charged current (CC) $ep\rightarrow\nu X$ in deep inelastic scattering
(DIS), shown in figures \ref{fig:WProdDiag}b and c respectively.
These processes are simulated by the RAPGAP \cite{RAPGAP} and DJANGO
\cite{DJANGO} generators. The regime of DIS is defined by the negative
momentum transfer square $Q^{2}$ of the boson exchanged between the
indicent electron and the quark scattered from the proton, at $Q^{2}>1$
GeV. The regime $Q^{2}<1$ GeV known as photoproduction is simulated
in the PYTHIA \cite{PYTHIA} framework. Here, two jets are produced
by boson-gluon-fusion as shown in figure \ref{fig:WProdDiag}d. For
the analysis presented in this paper NC and CC in DIS are important
background processes due to their high cross sections and event topologies.
They are also tools, because they provide high statistics samples
of final state objects such as leptons and QCD jets whose properties
need to be well understood to make accurate predictions. Figure \ref{fig:H1}b
shows measurements of NC and CC cross sections done at HERA.

As a template for physics beyond the Standard Model producing isolated
lepton and missing transverse momentum topologies, the anomalous production
of top quarks via flavor changing neutral currents (FCNC) is used
(shown in Fig. \ref{fig:WProdDiag}e). This process was generated
by the ANOTOP \cite{ANOTOP} programme and used to optimise acceptance
for a possible BSM signal.

\section{Search for Events with Isolated Electrons and Muons}

Electrons are detected as isolated, compact electromagnetic clusters
in the calorimeter and a high quality track pointing to that cluster.
Muons are detected by signals in the muon detectors, matching tracks
in the trackers and energy deposits in the LAr calorimeter typical
for minimally ionising particles. Electrons and Muons are detected
in the phase space of lepton transverse momentum $P_{T}^{l}>10$ GeV
and polar angular range $5^\circ\theta_{l}<140^\circ$ \cite{South03,Brandt07}.
Leptons are required to be isolated against other tracks and jets
in the event. This reduces leptons in or close to QCD jets which are
expected to be at lower $P_{T}^{l}$. In DIS the scattered electron
is mostly seen in the backward region $\theta_{l}>155^\circ$ of the detector.
The missing transverse momentum $P_{T}^{\mbox{miss}}$ is reconstructed
from the four-vector sum of all reconstructed final state particles
(hadrons and leptons) and required to be greater than 12 GeV. This
greatly reduces NC and photoproduction contributions to the background,
where $P_{T}^{\mbox{miss}}$ only appears due to fluctuations in the
energy measurement. Additionally SM processes are suppressed topologically,
for example by requiring acoplanarity of the lepton and the hadronic
final state four-vector in the transverse plane. For neutral current
and photoproduction events a back-to-back topology of the final state
objects is expected from momentum conservation.

\begin{figure}[t]
\begin{centering}\textbf{a)}\includegraphics[height=5.5cm]{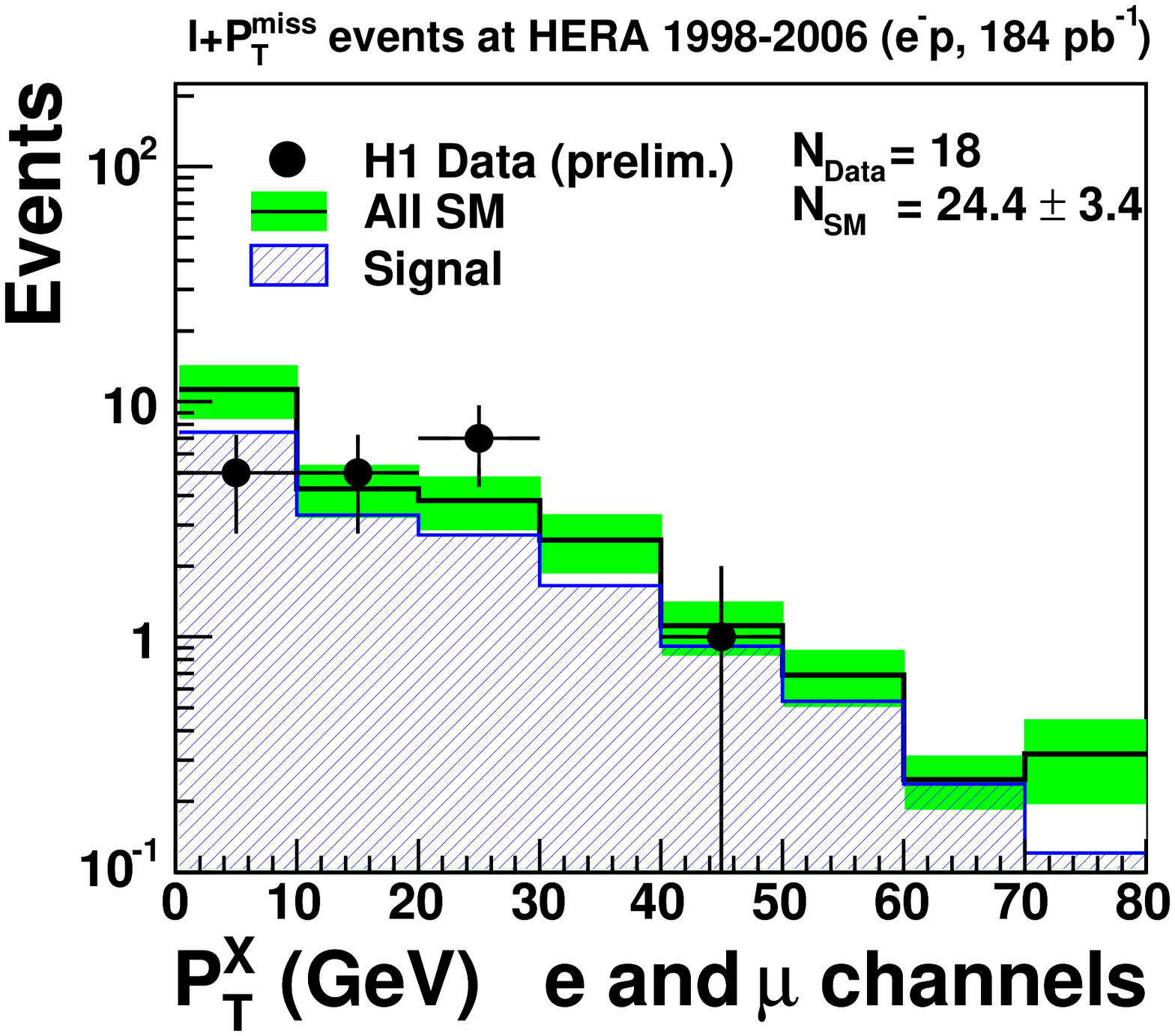}\textbf{b)}\includegraphics[height=5.5cm]{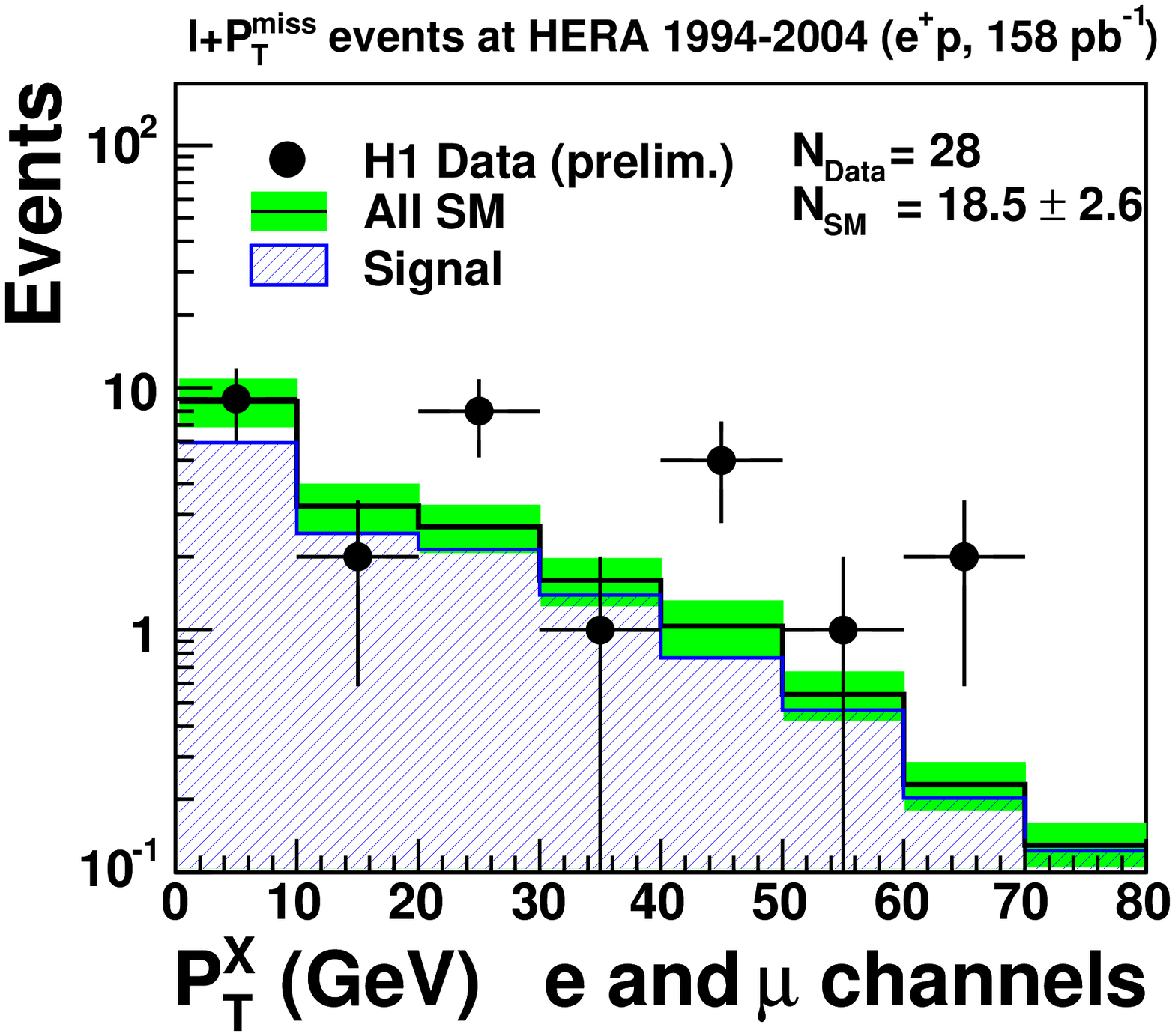} \par\end{centering}

\caption{\label{fig:elmuPTX}Transverse momentum of the hadronic system $P_{T}^{X}$
for the combined $e$- and $\mu-$channels in the $e/\mu+P_{T}^{\mbox{miss}}$
search. The points show the data with statistical errors. The open
histogram shows the SM expectation. Systematic and statistical errors
added in quadrature are shown as the shaded band. The signal contribution
of the SM expectation dominated by real $W$ production is shown as
a hatched histogram and amounts to about 70\%. While the $e^{-}p$
data (\textbf{a}) agree well with the SM expectation , the $e^{+}p$
data (\textbf{b}) show an excess of events over the SM prediction
amounting to 3.4$\sigma$ significance.}
\end{figure}

Figure \ref{fig:elmuPTX} shows the results of the search for the
combined electron and muon channels for the $e^{+}p$ data sample
corresponding to 158 pb$^{-1}$ and the $e^{-}p$ data sample corresponding
to 184 pb$^{-1}$. A total of 46 events are observed in the complete
data, with 43.0 $\pm$ 6.0 expected from the SM. At large hadronic
transverse momentum $P_{T}^{X}>25$ GeV a total of 18 events are observed,
but only 11.5 $\pm$ 1.8 predicted. 15 of these events are observed
in the $e^{+}p$ sample, where only 4.6 $\pm$ 0.8 are expected. Only
3 events are seen in the $e^{-}p$ data sample, where 6.9 $\pm$ 1.0
are expected. The significance of the observed excess in the $e^{+}p$
data sample is 3.4$\sigma$. No excess is observed in the $e^{-}p$
data sample. This observation represents a chance for discovery of
new physics at HERA.

\section{Search for Events with Isolated Tau Leptons}

Due to lepton universality in the Standard Model, the tau channel
is expected to behave identical to the electron and muon channels.
Leptonic $\tau$-decays are included in the electron and muon channels.
This search adds hadronic $\tau$-decays with one charged hadron,
leading to a 1-prong signature, which covers about 50\% of the total
$\tau$-lepton branching ratio \cite{Veelken06,Brandt07}. 3-prong
decays are not included because the background is too large to yield
a significant signal-to-background ratio.

Tau-jets are identified based on hadronic jets in the phase space
$P_{T}^{jet}>7$ GeV and $20^\circ<\theta^{jet}<120^\circ$. The candidate jets
are expected to be collimated due to the boost of the $\tau$-lepton.
This is ensured by requiring a jet radius $R_{jet}<0.12$ in the $\eta-\phi-$plane.
Exactly one track is required in a cone $D = 1.0$ in the $\eta-\phi-$plane 
around the
jet axis. Neutrinos from the decay chain $W\rightarrow\nu\tau\hookrightarrow l\nu_{l}\bar{\nu}$
are taken into account by requiring $P_{T}^{\mbox{miss}}>12$ GeV.
Because of the similarity of $\tau$-jets to narrow QCD jets with
a low track mulitplicity, the signal to background ratio is less favorable
than in the electron and muon channels. Jets in photoproduction ($Q^{2}<1$
GeV) may enter the sample due to fake $P_{T}^{\mbox{\mbox{miss}}}$
from fluctuations in the energy measurement and QCD jets faking $\tau$-jets.
CC background is much higher than in the electron and muon channel.
The correct description of dijet CC events becomes important since
their signature is very similar to the events in the interesting region
at high PTX. The second jet, which could fake a $\tau$-jet, may come
from next to leading order processes or final state radiation.

\begin{figure}[t]
\begin{centering}\textbf{a)}\includegraphics[height=5.5cm]{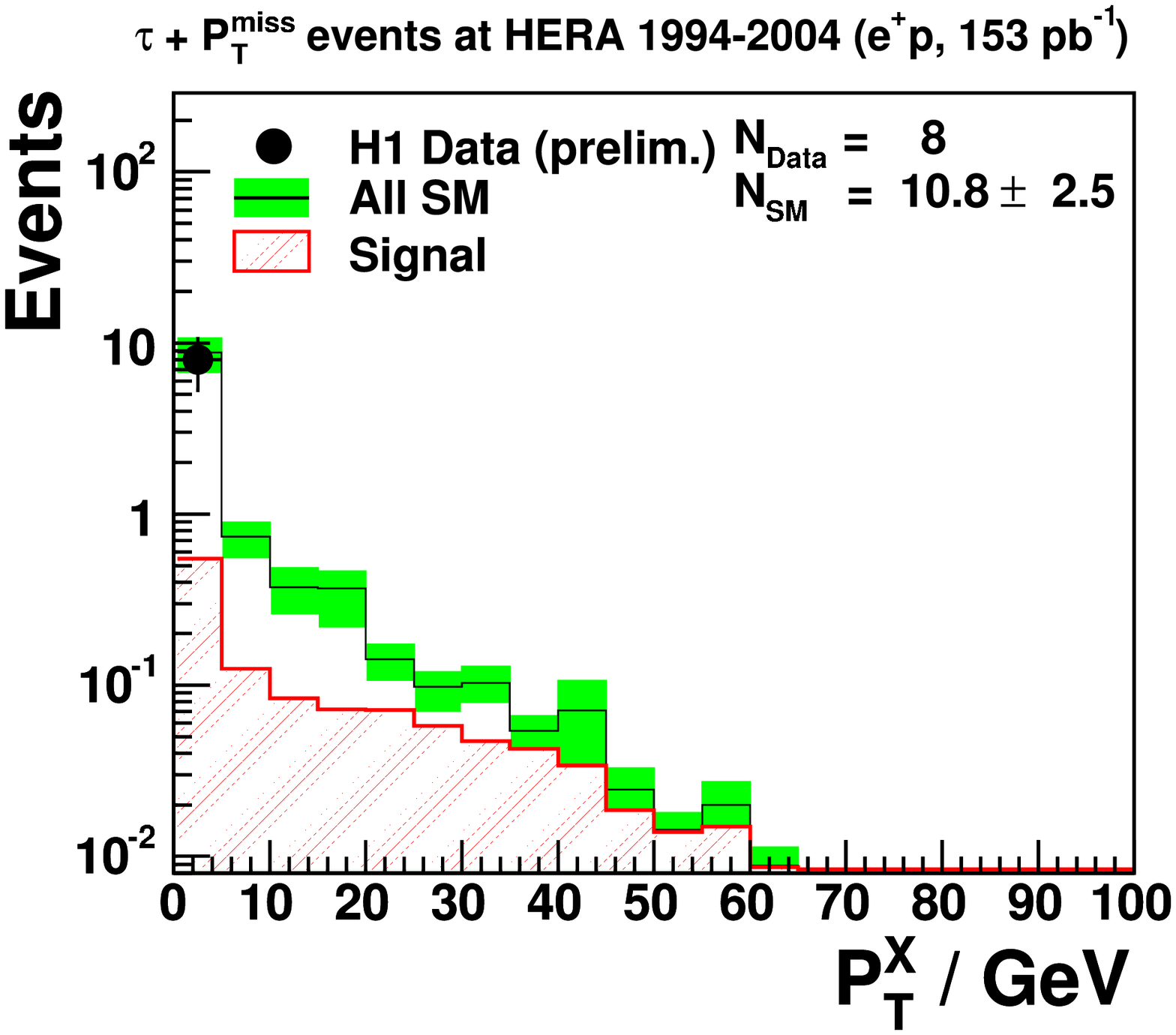}\textbf{b)}\includegraphics[height=5.5cm]{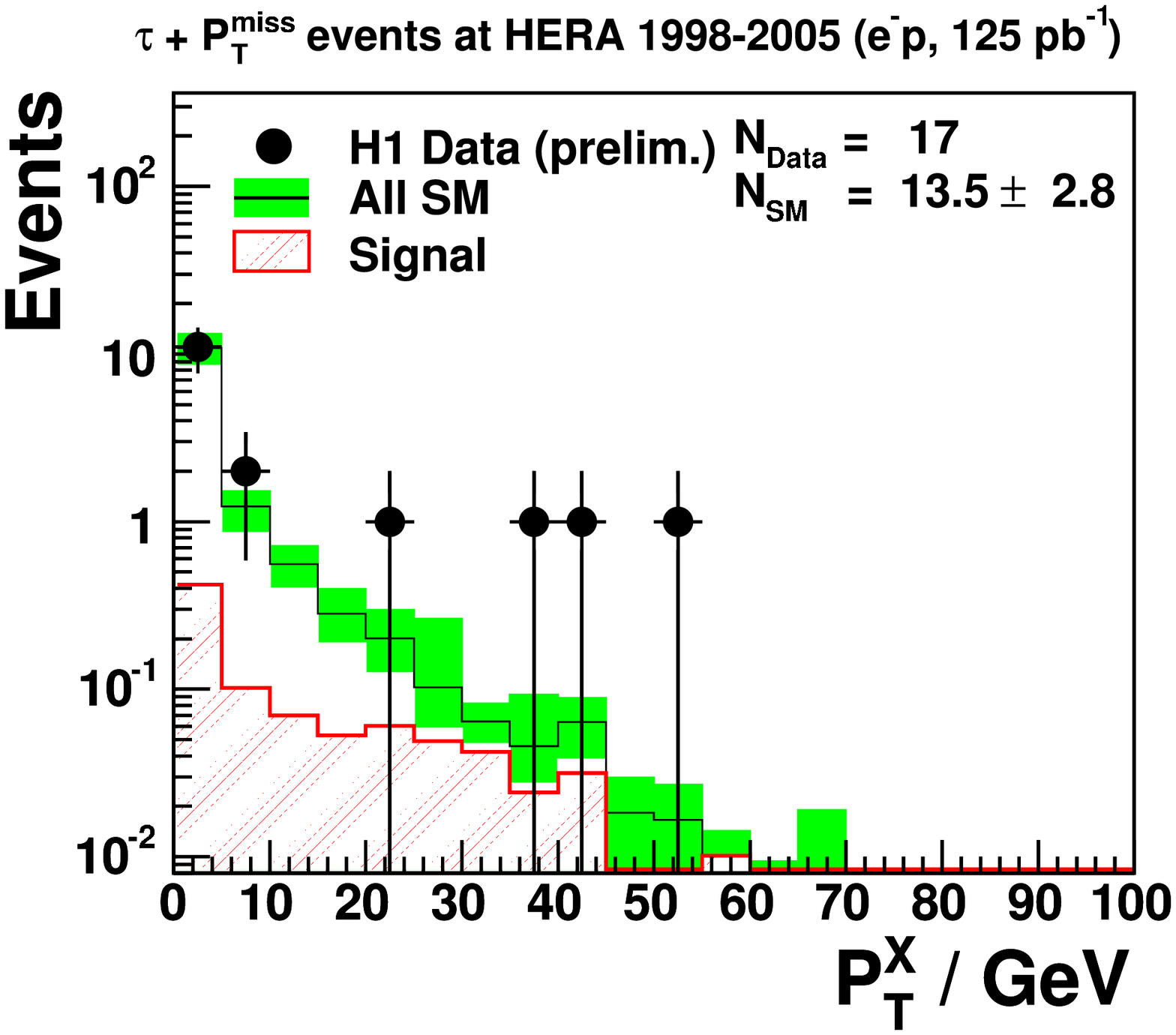}\par\end{centering}

\caption{\label{fig:TauRes}Transverse momentum of the hadronic system $P_{T}^{X}$
in $\tau+P_{T}^{\mbox{miss}}$ events. While there are no candidates
at high $P_{T}^{X}$ observed in the $e^{+}p$ sample (a), there are
three candidates at high $P_{T}^{X}$ observed in the $e^{-}p$ sample.}
\end{figure}

Figure \ref{fig:TauRes} shows the preliminary results for the $\tau$
analysis. In $e^{+}p$ data, 8 candidates are observed for 10.8 $\pm$
2.5 expected. All candidates are in the lowest $P_{T}^{X}$ bin where
most of the background is expected. In $e^{-}p$ data 17 events are
observed, with 13.5 $\pm$ 2.8 expected. Three events are observed
at $P_{T}^{X}$ > 25 GeV, while only 0.74 $\pm$0.18 are expected.
At first glance this seems to suggest a trend opposite to the electron
and muon channels, but this conclusion is premature. The statistical
significance of the result is low. Also at closer inspection, some
of the three observed candidates seem quite untypical for $\tau$-jets
and are more likely background that can possibly be removed in an
improved analysis.

\section{Interpretation}

While the low number of observed events in these analyses does not
yet show a clear pattern pointing to a strong hypothesis for a BSM
process to explain the excess, it is already intriguing to think about
possible mechanisms.

Upon the first observation of these events at HERA-1, one of the most
exciting interpretations was the production of single $top$-quarks.
While this is allowed kinematically and possible in CC processes at
HERA, the cross section is too low to see a signal in the available
data samples. If observed, $top$-quark events could therefore only
be produced anomalously via flavor changing neutral currents (FCNC).
The process is illustrated in figure \ref{fig:WProdDiag}e where the
scattered quark couples to the exchanged photon (or $Z^{0}$, not
shown) via a FCNC coupling and produces a $top$-quark. The subsequent
decay $t\rightarrow bW$ followed by $W\rightarrow l\nu$ or $W\rightarrow q\bar{q}$
produces either the searched $l+P_{T}^{Miss}$ signature or a 3 jet
topology. The $b$-jet gives rise to a highly energetic hadronic system
X, giving rise to the high $P_{T}^{X}$. These leptonic and hadronic
signatures have been analysed and limits have been set which are competitive
to limits on anomalous top production set at other colliders \cite{AnoTop03}.

Another possibility to explain the observation is to explicitly construct
a model. This could be based on a new particle $N$ with fermion number
F=0 coupling to $e-q$. If such a particle had a high mass $m_{N}$
it would have to be produced at large $x_{Bj}=M_{N}^{2}/s$. In this
region the parton density of valence quarks in the proton is much
larger than the density of sea quarks, hence the production cross
section for such a particle is much larger in $e^{+}p$ running than
in $e^{-}p$ running. In the framework of $R$-parity violating supersymmetry,
such particles could be resonantly produced $stop$-squarks, or $sbottom$-squarks
produced in the t-channel \cite{SUSY}. These preliminary ideas have
yet to be intensively studied.

A major issue with the observation is that ZEUS cannot confirm the
excess observed by H1, although the analysis is performed in a more
restricted phase space. Studies to properly compare the results and
test the compatibility of both experiments are ongoing \cite{IsoLepICHEP06}.

For the remainder of HERA data taking $e^{+}p$ data will be recorded.
If the data taking continues at the current rate, an additional integrated
luminosity of $\mathcal{L}\sim150$ pb$^{-1}$ can be expected. At
the time of writing the analysis of events with isolated leptons and
missing transverse momentum is followed with full effort to extract
the maximum of information from this puzzling observation.

{\footnotesize \lyxline{\footnotesize}}{\footnotesize \par}

{\footnotesize $^{3}$ H1 preliminary results available at:}{\footnotesize \par}

\texttt{\footnotesize http://www-h1.desy.de/publications/H1preliminary.short\_list.html}{\footnotesize \par}

{\footnotesize $^{4}$ ZEUS preliminary results available at:}{\footnotesize \par}

\texttt{\footnotesize http://www-zeus.desy.de}{\footnotesize \par}

{\footnotesize $^{5}$ H1 theses available at:}{\footnotesize \par}

\texttt{\footnotesize http://www-h1.desy.de/publications/theses\_list.html}
\end{document}